\documentclass[a4paper,11pt]{article}
\usepackage[table,xcdraw]{xcolor}
\usepackage{amsmath}
\usepackage{subcaption}
\usepackage{amsfonts}
\usepackage{amssymb}
\usepackage{authblk}
\usepackage{siunitx}
\usepackage{verbatim}
\usepackage[hidelinks]{hyperref}
\usepackage{hyperref} 
\usepackage{float}
\usepackage{graphicx} 
\usepackage[margin=25mm]{geometry}
\usepackage[numbers,sort&compress]{natbib}
\allowdisplaybreaks[4]
\title{Does the stability of f(R) theories imply the stability of the dual scalar-tensor theory?}

\author{Subham Chakraborty\thanks{\href{subchak001@gmail.com}{subchak001@gmail.com}}$^1$}
\author{Soumitra~SenGupta\thanks{\href{soumitraiacs1@gmail.com}{soumitraiacs1@gmail.com}}$^1$} 

\affil{$^1$ School of Physical Sciences, Indian Association for the Cultivation of Science, 2A $\&$ 2B Raja S.C. Mullick Road Kolkata - 700 032, India}

\begin{document}
\maketitle

\begin{abstract}
      Higher curvature f(R)  gravity theories are often plagued with Ostragadsky instability. In this work we show that such instability manifests itself in the corresponding dual scalar tensor theory in the scalar sector Lagrangian. We explicitly demonstrate the correspondence between the instabilities that appear in an $f(R)$ model and its corresponding scalar tensor theory. Considering various forms of f(R) gravity this feature is illustrated for different choices of the parameters of the theory.
      \end{abstract}

\section{Introduction}
The action of gravity  requires to have diffeomorphism invariance and the simplest choice for this is the Ricci scalar \cite{norton1993general}. However, it is also theoretically possible to use some function of the Ricci scalar, R, as the Lagrangian density. This leads to the concept of f(R) gravity which is considered to be one of the significant members among various classes of  higher curvature gravity theories \cite{capozziello2011extended,sotiriou2010f,nojiri2007introduction,sotiriou20096+,kalvakota2021investigating,berry2011linearized}. This theory gained significant attention through Starobinsky's work on cosmic inflation \cite{starobinsky1980new}. By choosing  an appropriate function of the Ricci scalar, f(R) gravity offers the potential to explain the accelerated expansion of the Universe and the formation of its structure without any need to invoke dark energy \cite{nojiri2011unified}. Such higher curvature action may have it's origin from  corrections arising from quantum gravity \cite{capozziello2011extended}, making it a relevant and natural  extension of general relativity. Though any correction of Einstein's action with $R^n$ ( $n \geq 2 )$ is  suppressed by inverse powers of Planck scale and  is therefore small compared to $R$ in the present regime, for epoch of high value of the curvature such corrections would definitely play crucial roles.
It is well-know that  f(R) theory introduces terms containing  higher derivatives of the metric leading to the appearance of unwanted Ostragadsky instability \cite{woodard2015theorem}. By appropriately choosing the $f(R)$ model one can get rid of such instabilities. A salient feature of $f(R)$ gravity theory is that by suitable conformal mapping of the metric function a higher curvature $f(R)$ gravity model can be cast into an equivalent scalar tensor theory where the degrees of freedom from higher curvature is encapsulated in a scalar field with a suitable potential term for the scalar \cite{fujii2003scalar}. The form of the scalar potential of course depends on the structure of the underlying $f(R)$ model.
The corresponding action of a scalar in the background of Einstein's gravity can be explored to study various cosmological/Astrophysical and other phenomenological features of the model without any complexities of the original action with higher derivative terms. However due to conformal transformation of the metric, the original $f(R)$ model and the dual scalar tensor theory may yield physically inequivalent consequences.  In this work we explore whether there exists any equivalence between the two models in the context of stability issues of the original $f(R)$ model and that in the corresponding scalar tensor theory. In other words, how does the instability of the original $f(R)$ originating from the higher derivative terms manifests itself in the scalar sector in Einstein's gravity? It there any one to one correspondence?
 This work reveals that by examining the properties of the potential of the scalar field as well as the nature of the kinetic term, it is possible to gain insights into the stability of the corresponding f(R) theory. This approach simplifies the stability analysis and provides a useful framework for understanding the dynamics and implications of various f(R) gravity models.

\section{The  actions for the modified gravitational
theory}

General 4-dimensional $f(R)$ action is given by:
\begin{equation} \label{a}
    S=\frac{1}{\kappa^2}\int d^4 x \sqrt{-g} f(R)\ .
\end{equation}
Here $R$ is the scalar curvature and $f(R)$ is some arbitrary function of $R$.

Such a  modification of the action from the original Einstein-Hilbert action leads to a new  set of field equations for the metric.

Introducing the auxiliary fields $A$ and $B$, one may rewrite the
action as follows:
\begin{equation} \label{b}
    S=\frac{1}{\kappa^2}\int d^4 x \sqrt{-g} \left\{B\left(R-A\right) + f(A)\right\}\ .
\end{equation}
By the variation over $B$one obtains  $A=R$. Substituting this into
(\ref{b}), 
the action (\ref{a}) can be reproduced. Now varying the action  with
 respect to $A$, we obtain
 
\begin{equation}\label{c}
    B=f'(A)
\end{equation}

Now one may eliminate $B$ by using (\ref{c}) and obtain
\begin{equation}\label{d}
S=\frac{1}{\kappa^2}\int d^4 x \sqrt{-g} \left\{f'(A)\left(R-A\right) + f(A)\right\}\ .
\end{equation}
At least classically, the two actions (\ref{d}) and (\ref{a}) are equivalent 
to each other \cite{nojiri2003modified}.

The action(\ref{d}) is  the Jordan
frame action with an  auxiliary field.
The corresponding Einstein frame action now may be obtained by suitable conformal transformation of the metric \cite{flanagan2004conformal} which we now demonstrate in the following section.

For the purpose of generality we derive here the  scalar-tensor action in Einstein's frame for the corresponding $f(R)$ model in arbitrary dimensional space-time.
The conformal transformation of the metric,
\begin{equation}
    g_{\mu\nu}\to e^\sigma g_{\mu\nu}\ ,
\end{equation}
leads to 
\begin{equation}
    \sqrt{-\tilde{g}} = e^{\sigma/2} \sqrt{-g}~.
\end{equation}

According to the transformation of the Ricci scalar, \cite{dabrowski2009conformal}

let,
\begin{equation}
\left(R^{(d)} - (d-1)\Box \sigma - {(d-1)(d-2) \over 4}
g^{\mu\nu}\partial_\mu \sigma \partial_\nu \sigma\right)\ = Z
\end{equation}

The modified action is now:

\begin{eqnarray*}
S^{d} &=& {1 \over \kappa^2}\int d^d x~ e^{\sigma/2} \sqrt{-g} \left\{f'(A) e^{-\sigma}Z-f'(A)A + f(A)\right\}\ \\
&=& \frac{1}{\kappa^2} \int d^d x~ [e^{\sigma d/2} ~ e^{-\sigma } f'(A)]\sqrt{-g} Z - {1 \over \kappa^2}\int d^d x~ e^{\sigma d/2} [f'(A)A-f(A)]\sqrt{-g}    \\
\end{eqnarray*}

Now, we chose the conformal factor in a way such that, 
\begin{equation}\label{eq:1}
    f'(A)= e^{\sigma}e^{-\sigma d/2}=e^{\frac{(2-d)\sigma}{2}}
\end{equation}
This by choice fixes the value of $f'(R)$ to be positive for all real values of $\sigma$ .
Also, inverting this we obtain,
\begin{equation}
    e^{\sigma d/2}=[f'(A)]^{-\frac{d}{d-2}}
\end{equation}

The whole action becomes:

\begin{eqnarray*}
S &=& {1 \over \kappa^2}\int d^d x~ \sqrt{-g} [R^{(d)} -(d-1)\Box \sigma - {(d-1)(d-2) \over 4} 
g^{\mu\nu}\partial_\mu \sigma \partial_\nu \sigma - V(\sigma)]\\
\end{eqnarray*}

where, 
\begin{equation}\label{+ve}
    {V(\sigma) = \frac{f'(A)A-f(A)}{(f'(A))^{\frac{d}{d-2}}}}
\end{equation}

$(d-1)\Box \sigma$ term would contribute to the boundary term, hence can be ignored.
So the action for scalar tensor theory of gravity in Einstein frame in $d$ dimension is given by:

\begin{equation}
S^{d} = \frac{1}{\kappa^2} \int d^d x~ \sqrt{-g} [R^{(d)} - {(d-1)(d-2) \over 4} g^{\mu\nu}\partial_\mu \sigma \partial_\nu \sigma - V(\sigma)]
\end{equation}

Thus the f(R) action transforms into a dual scalar tensor theory. The scalar $\sigma$ comes from the conformal transform of the metric tensor. 
The instability of the original $f(R)$ theory  has it's origin in the higher derivative terms in the $f(R)$ model. We want to explore whether it manifests itself now as an  instability in the scalar sector of the  corresponding scalar tensor theory which may appear from the signature of the kinetic term as well as the form of the `potential' $V(\sigma)$. The structure of the scalar potential in turn depends on the type of $f(R)$ model that is being considered. In the coming sections we are going to do an analysis of the stability of the f(R) model well as the dual scalar-sector model systematically to establish the correspondence between them.

In contrast to Eq. (\ref{eq:1}) when we chose $f'(R)>0$, we consider $f'(R)$ as follows,
\begin{equation}\label{-ve}
    f'(A)= - e^{\sigma}e^{-\sigma d/2}= - e^{\frac{(2-d)\sigma}{2}}
\end{equation}
This by choice fixes the value of $f'(R)$ to be negative.
Also, inverting this,
\begin{equation}
    e^{\sigma d/2}=[-f'(A)]^{-\frac{d}{d-2}}
\end{equation}
Similar algebra leads to the final action in this case.
\begin{equation}\label{f'r}
{S = \frac{1}{\kappa^2} \int d^d x~ \sqrt{-g} [-R + {(d-1)(d-2) \over 4} g^{\mu\nu}\partial_\mu \sigma \partial_\nu \sigma - V(\sigma)]}
\end{equation}

where, 
\begin{equation}\label{-vs}
    {V(\sigma) = \frac{f'(A)A-f(A)}{(-f'(A))^{\frac{d}{d-2}}}}
\end{equation}

In our present sign convention (Minkowski metric being $(-,+,+,+)$) from Eq.(\ref{f'r}) the sign of kinetic term of the scalar implies that the system is unbounded from below. So, $f'(R)>0$ becomes an essential condition to get rid of the instability originating from the kinetic term of the scalar sector. 

\section{A General approach for positive $f'(R)$}
Without choosing any particular form of f(R), we first try to examine the stability issue as generally as possible for (3+1) dimensional space-time.
We found a relation between the scalar potential and $f(A)$ in (\ref{+ve}). We presume this scalar potential must be real, should have at least one minima and should be bounded from below to ensure stability. 
Now, in order to find a minima in that potential, we obtain,
\begin{equation}
    \frac{d V(\sigma)}{d \sigma}=\frac{d V(\sigma)}{d(f'(A))}\frac{d (f'(A))}{d \sigma}
\end{equation}
and, 
\begin{equation}
    \frac{d^2 V(\sigma)}{d \sigma^2}= \frac{d}{d(f'(A))} \left(  \frac{d V(\sigma)}{d(f'(A))}\frac{d (f'(A))}{d \sigma}  \right) \frac{d (f'(A))}{d \sigma}
\end{equation}

Also, from the choice of conformal factor ( for d = 4), $f'(A)=e^{-\sigma}$. So, 
\begin{equation}
    \frac{d (f'(A))}{d \sigma}=- f'(A)
\end{equation}
From Equation (\ref{+ve}) we have,
\begin{equation}
    \frac{d V(\sigma)}{d(f'(A))}= \frac{A}{(f'(A))^2}+2\frac{f(A)}{(f'(A))^3}
\end{equation}
Putting this in the  above equations, we can get the conditions for extrema for the scalar potential by setting first derivative with respect to the scalar $\sigma$ to zero and obtain,
\begin{equation}
    Af'(A)=2f(A)
\end{equation}
For the extrema to be minima we must have $\frac{d^2 V(\sigma)}{d \sigma^2}>0$, which gives,
\begin{equation}
    \frac{1}{f''(A)}>\frac{A}{f'(A)}
\end{equation}
Now the sign of $f''(A)$ depends on sign of $A$ at the minima. In other words, to ensure a minima, $f''(A)$ can be both positive or negative depending on the exact sign of A at minima. This feature will be further clarified in later sections where we will consider specific choices of $f(R)$ model.

\section{$f(R)=R+\alpha R^n$ }

Here $R$ can be expressed in terms of  the scalar $\sigma$ explicitly by inverting the relation between them and hence the conditions on the parameters namely (n, $\alpha$) can be derived in order to get a stable theory. 
For sake of generality, we obtain the expression in arbitrary dimension $d$.  

For $f(R)=R+ \alpha R^n$, we find,
\begin{eqnarray*}
    1+ \alpha nR^{n-1}&=& e^{\frac{(2-d)\sigma}{2}}
\end{eqnarray*}

from which,

\begin{eqnarray*}
     R &=& \left[ \frac{1}{ \alpha n} (e^{\frac{-(d-2)\sigma}{2}} -1)  \right]^{\frac{1}{n-1}}
\end{eqnarray*}

From the form of $V(\sigma)$ in terms of f(R), and putting the value of R from the above equation.

\begin{eqnarray}\label{2}
   { V(\sigma)=\left[\frac{ \alpha (n-1)}{( \alpha n)^{\frac{n}{n-1}}}\right] \cdot  \frac{\left(  e^{\frac{(2-d)\sigma}{2}} -1  \right)^{\frac{n}{n-1}}}{e^{-\frac{\sigma d}{2}}} } 
\end{eqnarray}

The final scalar potential for $f(R)=R+ \alpha R^n$ in a spacetime dimension of d is given in (\ref{2})

As we demanded previously that the f(R) should be stable if the corresponding scalar potential $V(\sigma)$ is real and has at least one minima for some value of $\sigma$ and finally it should be bounded from below, thus should not approach negative infinity for some value of $\sigma$.

So now we can find the extrema for the potential and  note all cases for which at least one minima is observed. Also for the rest the analysis is only for 4d spacetime.

\section{Stablility analysis for odd value of $n$}

The corresponding potential for odd values of $n=2m+1$ is given by
\begin{equation}
V(\sigma) = (2m) \alpha \left[  \frac{e^{-\sigma} - 1}{\alpha (2m+1)}  
 \right]^{\frac{2m+1}{2m}} e^{2\sigma}~~~~~~~~~~~~~~~~~~~~~~~~~~~~~~~~~n=2m+1~~,\forall m \in \mathbb{N}
\end{equation}

We note ($e^{-\sigma} - 1$) is negative for $\sigma >0$ and Positive otherwise. Also the even value in the denominator of power leads the potential to be imaginary on one side of $\sigma =0$
 depending on the sign of $\alpha$. Clearly, for positive $\alpha$ the potential is imaginary in case of $\sigma>0$ and for negative $\alpha $ the potential is imaginary for $\sigma<0$. To illustrate these properties $\alpha=+1$ or, $-1$ are taken in the case of $n=3$ below as examples.
\subsection{Positive value of $\alpha$ {\\$n=3~~~~ \alpha =+1~~~$\\ } }

The form of the corresponding potential in terms of $\sigma$:
\begin{equation}
V(\sigma) = 2 \left(\frac{e^{-\sigma} - 1}{3}\right)^{\frac{3}{2}} e^{2\sigma}
\end{equation}
Potential V is only real for negative $\sigma$. \\
        {Also only one maxima is found in $\sigma$. Hence, the f(R) theory shows scalar instability for $\alpha >0$ and $n=3$.}

\subsection{Negative value of $\alpha$ {\\$n=3~~~~ \alpha =-1~~~$\\ }}

The form of the corresponding potential in terms of $\sigma$:
\begin{equation}
V(\sigma) = -2 \left(\frac{e^{-\sigma} - 1}{-3}\right)^{\frac{3}{2}} e^{2\sigma}
\end{equation}
Also it does not attain any minima in the region it is real. Hence, the f(R) theory shows scalar instability for $\alpha < 0$ and $n=3$.

\begin{figure}[H]
\centering
    \begin{subfigure}{0.45\textwidth}
        \includegraphics[scale=0.55]{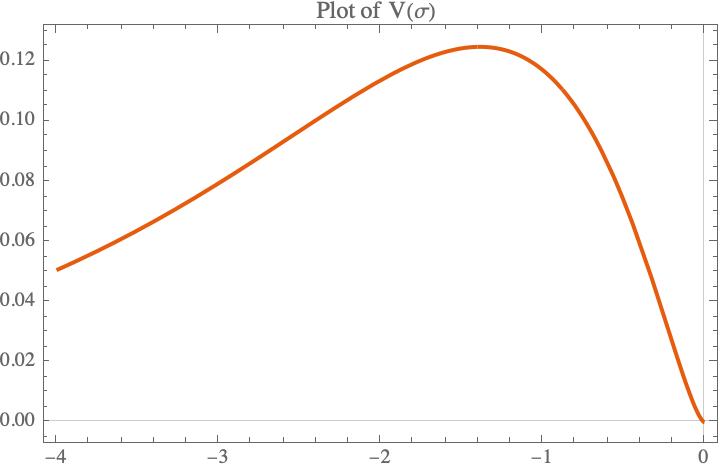}
        \caption{{ $\alpha$ =$1$}}
        \label{fig:enter-label}
    \end{subfigure}
    \begin{subfigure}{0.45\textwidth}
        \includegraphics[scale=0.55]{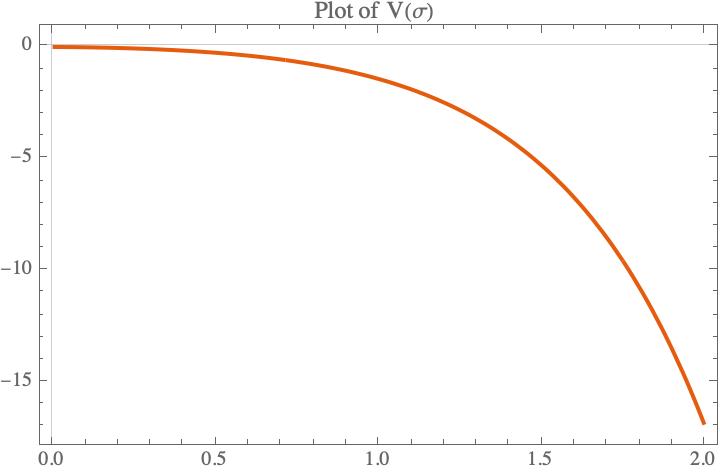}
        \caption{{$\alpha$ =$-1$}}
        \label{fig:enter-label}
    \end{subfigure}
    \caption{Plot of the scalar potential in 4 dimension, regarding odd values of $n$. Here $n=3$.  }
    \label{fig:enter-label}
\end{figure}

As mentioned before we deduce that $n=3$ and other odd numbers, {\textit{the potential sector does not show stability for either case $\alpha <0$~~or~~$\alpha>0$}}.

\section{Stablility analysis for Starobinsky model ($n=2$)}
Corresponding potential is given by,
\begin{equation}
V(\sigma) =  \alpha \left(\frac{e^{-\sigma} - 1}{2 \alpha}\right)^{2} e^{2\sigma}
\end{equation}

Clearly the character of the potential directly depends on the sign of $\alpha$. The $\alpha$
 outside the braces if positive, ensures the potential to attain a minima at $\sigma=0$ and if negative, ensures the potential to attain a maxima at $\sigma=0$. The plots of these are shown below with $\alpha =1 $and$ -1$
\subsection{Positive value of $\alpha$ {\\$n=2~~~~ \alpha =+1~~~$\\ }}

The form of the corresponding potential in terms of $\sigma$:
\begin{equation}
V(\sigma) = \left(\frac{e^{-\sigma} - 1}{2}\right)^{2} e^{2\sigma}
\end{equation}
Only one minima is found in $\sigma=0$. Hence, the f(R) theory shows stability for $\alpha > 0$ and $n=2$. This is exactly the solution given by \textit{Starobinsky} \cite{starobinsky1980new}. Starobinsky noted that quantum corrections to general relativity should be important for the early universe. These generically lead to curvature-squared corrections to the Einstein–Hilbert action and a form of f(R) modified gravity \cite{starobinskii1979spectrum,vilenkin1985classical}. The solution to Einstein's equations in the presence of curvature squared terms, when the curvatures are large, leads to an effective cosmological constant \cite{cembranos2009dark}. Therefore, he proposed that the early universe went through an inflationary de Sitter era \cite{starobinskii1983perturbation}. This resolved the cosmology problems and led to specific predictions for the corrections to the microwave background radiation, corrections that were then calculated in detail.

\subsection{Negative value of $\alpha$ {\\$n=2~~~~ \alpha =-1~~~$\\ }}

The form of the corresponding potential in terms of $\sigma$:
\begin{equation}
V(\sigma) = -\left(\frac{e^{-\sigma} - 1}{-2}\right)^{2} e^{2\sigma}
\end{equation}
Therefore only one maxima is found in $\sigma=0$. Hence, the f(R) theory doesn't show stability for $\alpha < 0$ and $n=2$.

\begin{figure}[H]
\centering
    \begin{subfigure}{0.45\textwidth}
        \includegraphics[scale=0.55]{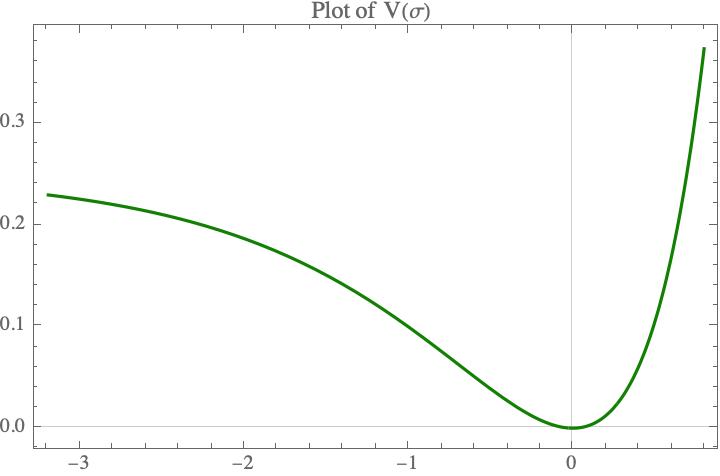}
        \caption{$\alpha$ =$1$}
        \label{fig:enter-label}
    \end{subfigure}
    \begin{subfigure}{0.45\textwidth}
        \includegraphics[scale=0.55]{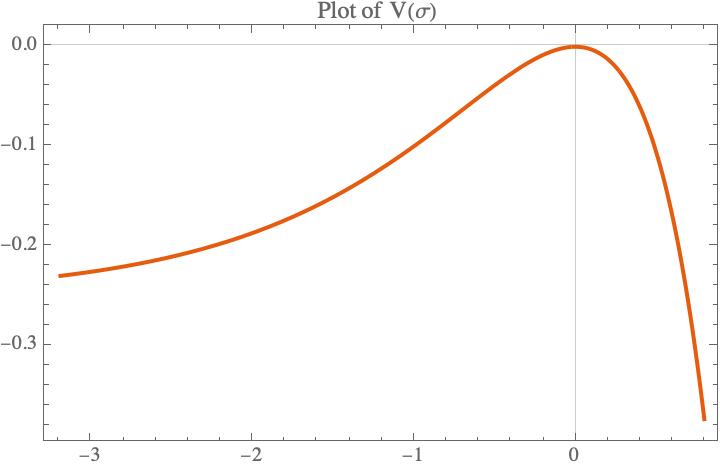}
        \caption{$\alpha$ =$-1$}
        \label{fig:enter-label}
    \end{subfigure}
    \caption{Plot of the scalar potential in 4 dimension, regarding $n=2$.  }
    \label{fig:enter-label}
\end{figure}

\section{Stablility analysis for $n$ even other than $2$}

Here we take $n$ to be any even integer($2m$) other than 2. Corresponding potential is given by,
\begin{equation}
V(\sigma) =  (2m-1) \alpha \left[ \frac{e^{-\sigma} - 1}{2m \alpha}\right]^{\frac{2m}{2m-1}} e^{2\sigma} ~~~~~~~~~~~~~~where,~~m=2, 3, 4,...
\end{equation}

Clearly the character of the potential directly depends on the sign of $\alpha$. The sign of $\left[ \frac{e^{-\sigma} - 1}{2m \alpha}\right]$ does not matter because of the even numerator in power. The $\alpha$
 outside the brackets if positive, ensures the potential to attain a minima at $\sigma=0$ and a maxima at some negative $\sigma$ and if negative, ensures the potential to attain a maxima at $\sigma=0$ and a minima at some negative $\sigma$. 
To illustrate these properties $\alpha=+1$ or, $-1$ are taken in the case of $n=4$ (i.e. $m=2$) below as examples.

\subsection{Positive value of $\alpha$ {\\$n=4~~~~ \alpha =+1~~~$\\ }}

The form of the corresponding potential in terms of $\sigma$:
\begin{equation}
V(\sigma) = 3\left(\frac{e^{-\sigma} - 1}{4}\right)^{\frac{4}{3}} e^{2\sigma}
\end{equation}

Therefore only one minima is found in $\sigma=0$ and one maxima at $\sigma=-1.099$. Hence, the f(R) theory should show stability for $\alpha > 0$ and $n=4$.

\subsection{Negative value of $\alpha$ {\\$n=4~~~~ \alpha =-1~~~$\\ }}

The form of the corresponding potential in terms of $\sigma$:
\begin{equation}
V(\sigma) = -3\left(\frac{e^{-\sigma} - 1}{-4}\right)^{\frac{4}{3}} e^{2\sigma}
\end{equation}

The graph shows one maxima at $\sigma=0$ and one minima at $\sigma=-1.099$. Despite getting a minima at a non zero value of $\sigma$, the minima is meta-stable. Moreover, the potential being unbounded from below can not lead to a stable theory for $\alpha <0$ and $n=4$.

Here we refer our earlier discussion, where we showed, if there exists a minima, still one can find $f''(R)$ to be negative. Because $f(R)=R-R^4$ gives negative $f''(R)$ for any values of R.
\begin{figure}[H]
\centering
    \begin{subfigure}{0.45\textwidth}
        \includegraphics[scale=0.55]{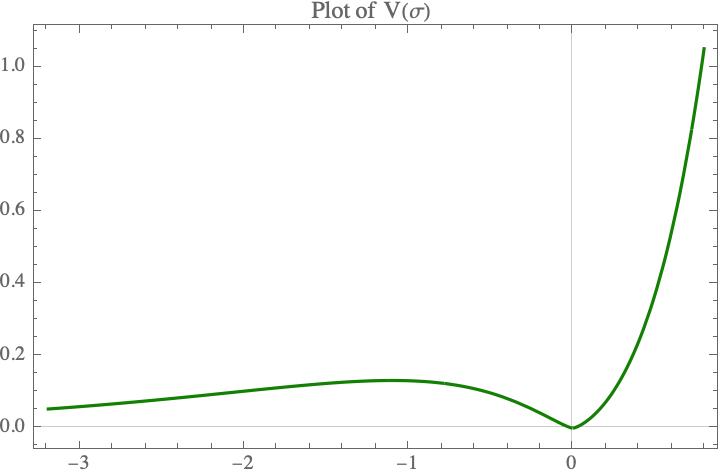}
        \caption{{$\alpha=1$}}
        \label{fig:enter-label}
    \end{subfigure}
    \begin{subfigure}{0.45\textwidth}
        \includegraphics[scale=0.55]{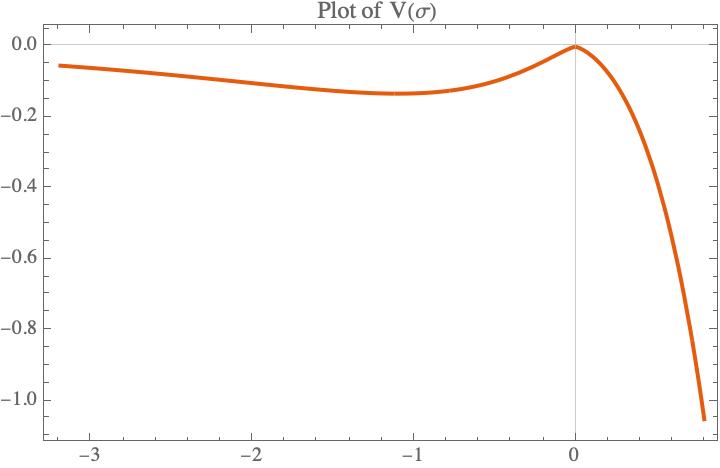}
        \caption{{$\alpha=-1$}}
        \label{fig:enter-label}
    \end{subfigure}
    \caption{Plot of the scalar potential in 4 dimension, regarding even values of $n$ except $n=2$. Here value of $n$ is taken to be $4$.  }
    \label{fig:3}
\end{figure}

\section{Generalisation}

We saw that $f(R)=R+\alpha R^n$ if analysed by the corresponding scalar doesn't give stability for all values of $\alpha$ and n.

The odd powers of R does not show stability in the scalar sector for any f(R). Also, any negative value of $\alpha$ leads to instability in scalar sector.

\begin{figure}[H]
    \centering
    \begin{minipage}{0.35\textwidth}
        \centering
        \includegraphics[scale=0.41]{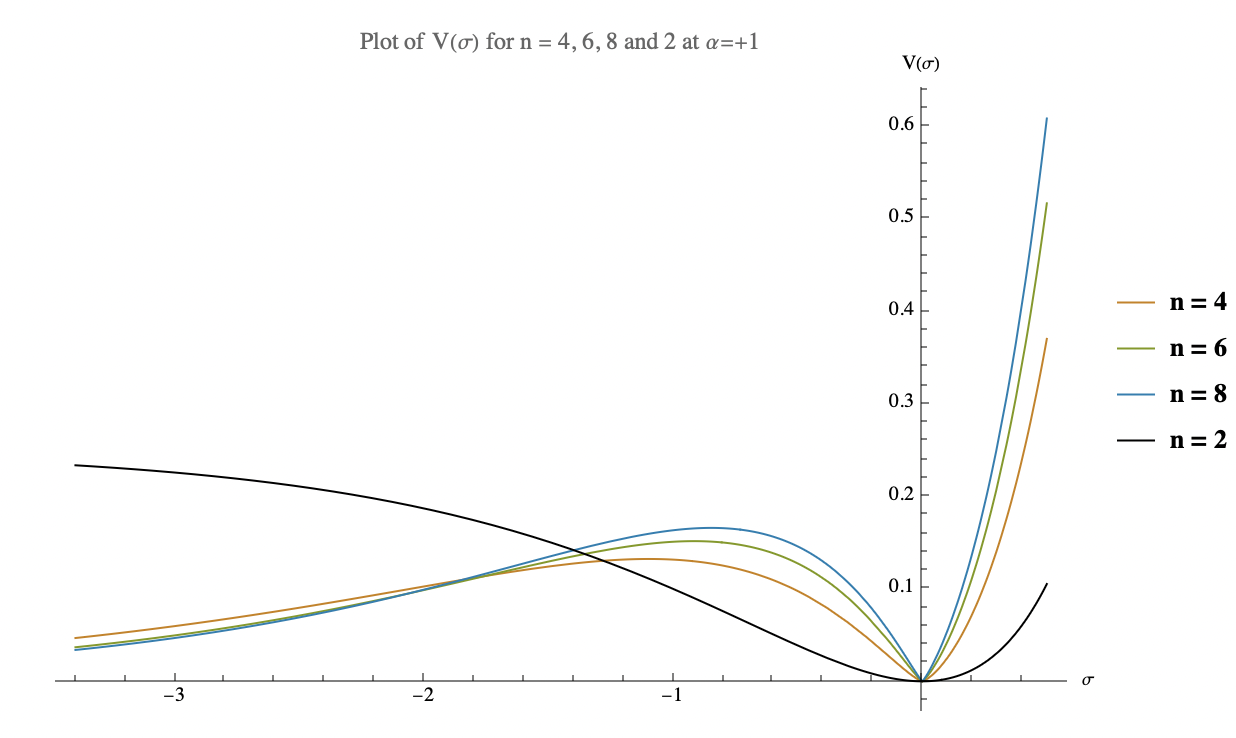}
        \caption{Minima at origin: \\even n for $\alpha=+1$ }
        \label{fig:enter-label}
    \end{minipage}\hspace{0.27\textwidth}
    \begin{minipage}{0.36\textwidth}
        {Through out the analysis, though some example parameters are considered to see the behaviour of the potential in each case, same analysis have been done for other parameters as well and the results are similar. Here is a parametric plot of potentials regarding positive $\alpha$ (+1), for different even $n$. Clearly every potential has at least one minima at $\sigma=0$}
    \end{minipage}
\end{figure}

\begin{figure}[H]
    \centering
    \begin{minipage}{0.35\textwidth}
        \centering
        \includegraphics[scale=0.47]{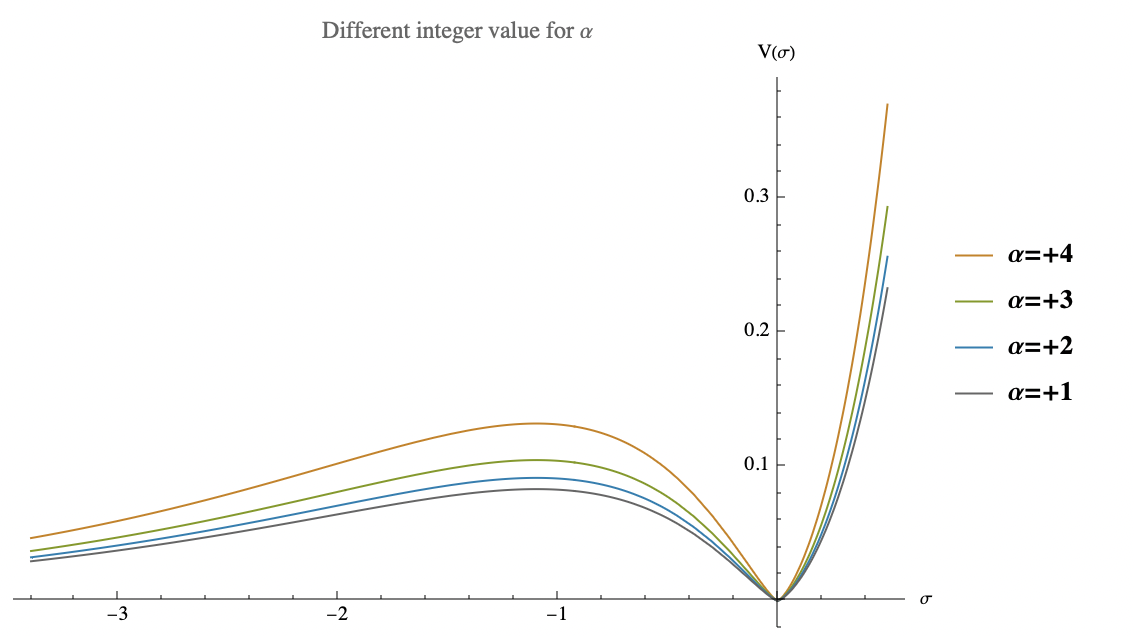}
        \caption{Minima at origin: \\ n=4 for $\alpha=+1, +2, +3, +4$ }
        \label{fig:enter-label}
    \end{minipage}\hspace{0.27\textwidth}
    \begin{minipage}{0.36\textwidth}
        {
Same analysis is done for different values of $\alpha$ as well. The nature of the potential remains exactly same again, with same number of minima or maxima at the origin. The position of additional maxima or minima however, changes. This is seen by plotting the potential for different positive integer values of $\alpha$. Also any value of negative $\alpha$ leads to instability.
}
    \end{minipage}
\end{figure}


For a higher curvature theory of the form $f(R)=R+\alpha R^n$ to be stable in scalar sector, $n$ must be even and $\alpha$, must be positive.

\section{Analysing $f(R)=R+\alpha R^n+\beta R^m$
}

The above analysis is then extended for one more term of $ \beta R^m$ in $f(R)$, where $m$ is again an integer. Taking specific combinations of odd and even values of $m$ and $n$, as well as taking positive and negative value of $\alpha$ and $\beta$ leads to several scalar potentials. 

Only scalar potentials $V(\sigma)$ regarding $f(R)=R+ |a| R^{2p} + |b| R^{2q}$ and $f(R)=R- |a| R^{2p} - |b| R^{2q}$ , ($p,q$ integer) turned out to be real for all values of $\sigma$. For all other cases, i.e. alternate sign signature of $\alpha$, $\beta$ and odd values of $m$ or $n$, the potentials were found out to be complex for different range of $\sigma$.

The real scalar potentials for $n=2$ and $m=4$ are plotted below.

\begin{figure}[H]
\centering
    \begin{subfigure}{0.45\textwidth}
        \includegraphics[scale=0.55]{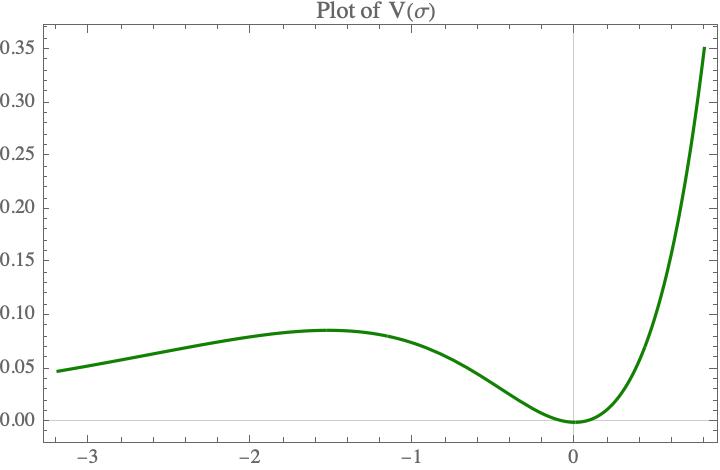}
        \caption{{$\alpha=+1,~\beta=+1$}}
        \label{fig:6a}
    \end{subfigure}
    \begin{subfigure}{0.45\textwidth}
        \includegraphics[scale=0.55]{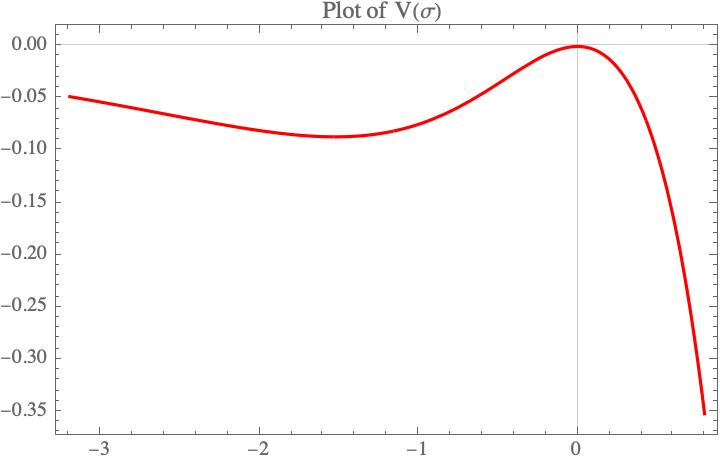}
        \caption{{$\alpha=-1,~\beta=-1$}}
        \label{fig:6b}
    \end{subfigure}
    \caption{Plot of the scalar potential in 4 dimension for $f(R)=R+\alpha R^n+\beta R^m$, regarding even values of $n$ and $m$. Here value of $n$ is $2$ and $m$ is $4$.  }
    \label{fig:enter-label}
\end{figure}

Clearly These shows similar characteristics as figure (\ref{fig:3}). figure (\ref{fig:6a}) has one minima at $\sigma=0$, hence is stable in scalar sector.  For figure (\ref{fig:6b}) on the other hand has a meta-stable minima at some non zero value of $\sigma $. But the potential is unbounded from below, so can not lead to a stable theory.

So for a higher curvature theory of the form $f(R)=R+\alpha R^n+\beta R^m$ to be stable in scalar sector, $m$, $n$ must be even and $\alpha$, $\beta$ must be positive.

Again, for simplicity we have considered upto 4th power of $R$ in $f(R)$ and modulus of $\alpha$, $\beta$ to be unity. But this is checked using higher powers in $R$ also for arbitrary value of $\alpha$, $\beta$ leading to same result.

\section{Correspondence between Ostrogradsky Instability and scalar sector instability }
Ostrogradsky instability regarding f(R) theory demands the theory can be stable in case of $f'(R)>0$ and $f''(R)>0$ \cite{de2016ostrogradsky}. 

From (\ref{f'r}) we eliminated the case of negative $f'(R)$ with an argument regarding the sign of kinetic term of the scalar field in the action. So, in scalar sector $f'(R)>0$ condition holds. 

Also if we consider the solutions of $f(R)$ for which the scalar sector shows stability, 
in case of $f(R)$ being in the form of $R+\alpha R^n$ :
\begin{eqnarray*}
    f(R)&=&R+|a|R^{2p}\\
    f''(R)&=& 2p (2p-1) |a|R^{2p-2}
\end{eqnarray*}

We note that,
$f''(R)$ is indeed positive for any integer $p$ or any value of R.

Now, for $f(R)=R+\alpha R^n+\beta R^m$, scalar sector stable solutions:
\begin{eqnarray*}
    f(R)&=&R+|a|R^{2p}+|b|R^{2q}\\
    f''(R)&=& 2p (2p-1) |a|R^{2p-2} + 2q (2q-1) |b|R^{2q-2}
\end{eqnarray*}
Again $f''(R)$ is positive for any integer $p$, $q$ or any value of R ($a$,$b$ $\in$ $\mathbb{R}$).

Conversely, $f''(R)$ is positive only for these solutions. i.e. for odd powers of $R$ and negative coefficients, $f''(R)<0$ for at least some range of values of $R$. 

\textit{Which indicates a one-to-one correspondence of scalar sector instability with Ostrogradsky instability of higher curvature theories.}
\section{Energy Conditions}

For a scalar field in action the stress–energy tensor is given by,

\begin{equation}
    T_{\alpha \beta}=\nabla_{\alpha} \phi \nabla_{\beta} \phi - \frac{1}{2} g_{\alpha \beta}(\nabla_{\mu} \nabla^{\mu} + V(\phi))
\end{equation}
Now to check weak energy condition we found, 
\begin{equation}
    T_{\alpha \beta} U^{\alpha}U^{\beta}=\frac{1}{2}(\nabla_0 \phi)^2 + \frac{1}{2}(\nabla_i \phi)^2+ \frac{1}{2} V(\phi) \geq 0
\end{equation}
where, $U^{\mu}$ is a time-like vector. Clearly positive potentials ensures weak energy condition is obeyed. Now for some cases we found scalar sector instability, the graph of scalar potentials showed, for those particular cases the scalar potential $V(\sigma)$ was unbounded from below, i.e. attained negative infinity at some point denoting the fact that for those cases, weak energy condition is violated. This gives another view of why we did not consider those cases to be stable despite having one minima (meta-stable) at some non-zero value of the potentials. 

\section{Conclusion}
In conclusion, our comprehensive analysis of the f(R) gravity framework provides significant insights into the stability and viability of these modified gravity theories. We began by establishing the action for f(R) gravity, which allowed us to introduce the Jordan frame action through the incorporation of auxiliary fields. This foundational step set the stage for a deeper exploration of the mathematical and physical implications of f(R) gravity. By employing a conformal transformation, we were able to convert the action into a scalar-tensor form, which is a critical step in simplifying the complex equations that govern these theories. This transformation enabled us to derive a fundamental relationship between the scalar potential and f(R). 
Our study then focused on two specific forms of f(R), namely $f(R)=R+\alpha R^ 
n$ and, $f(R)=R+\alpha R^ n+ \beta R^m$
 , which provided a concrete example to investigate the dynamics and stability of the potential. Through detailed  analysis, we examined the conditions under which the scalar potential is real, exhibits at least one minima and bounded from below. This finding is pivotal as it outlines the fundamental criteria that the dual scalar tensor model must satisfy to be considered stable. Since the scalar potential inherits the parameters $\alpha$ and $\beta$ from the original structure of the $f(R)$
 theory, we detailed the constraints on the parameters $\alpha, n, \beta, m $. We saw, $\alpha, \beta$ must be positive and $m,n$ must be even for this to happen. Now, these are the only cases where $f''(R)$ is positive for all values of $R$ 
 and $f'(R)$ is already positive to keep the correct signature of the kinetic term of the scalar field in action. 
Hence, we highlighted the correspondence between Ostrogradsky instability and the scalar potential instability. Our result reveals that the stability of one implies the stability of the other.


\bibliographystyle{unsrt}
\bibliography{ref}

\end{document}